\documentclass[12pt]{article}
\usepackage{geometry}
\usepackage{enumitem,amssymb}
\usepackage{ragged2e}
\newlist{thematic}{itemize}{8}
\setlist[thematic]{label=$\square$}
\usepackage{pifont}
\newcommand{\cmark}{\ding{51}}%
\newcommand{\done}{\rlap{$\square$}{\raisebox{2pt}{\large\hspace{1pt}\cmark}}%
\hspace{-2.5pt}}

\usepackage{booktabs} 
\usepackage{array} 
\usepackage{verbatim} 
\usepackage{graphicx}
\usepackage[colorlinks,
            linkcolor=blue,
            citecolor=blue,
            filecolor=blue,
            urlcolor=blue]{hyperref}
\usepackage{tabularx} 
\usepackage[table]{xcolor} 
\usepackage{wasysym} 
\usepackage{subfigure} 

\usepackage{enumitem}
\usepackage{lineno}
\usepackage[numbers,sort&compress]{natbib}

\begin{document}
\thispagestyle{empty}
\huge
\noindent Astro2020 Science White Paper \\[0.5cm]
\noindent The Sun at GeV--TeV Energies:\\ A New Laboratory for Astroparticle Physics\\

\normalsize

 \noindent \textbf{Thematic Areas:} \done \hspace{2pt}  Multi-Messenger Astronomy and Astrophysics.\\ \done  \hspace{2pt} Cosmology and Fundamental Physics. \hspace*{10pt} \done \hspace{2pt}  Stars and Stellar Evolution.
 \\

\setcounter{page}{0}

\noindent \textbf{Principal Author:}

\noindent Name: Mehr Un Nisa 
					
\noindent Institution: University of Rochester  

\noindent Email: \href{mailto:}{mehr.unnisa@icecube.wisc.edu}\hspace*{20pt}Phone: +1 (857) 204 8765\\

\noindent\textbf{Co-authors:} \href{mailto:beacom.7@osu.edu}{J. F. Beacom} (The Ohio State University), \href{mailto:sbenzvi@ur.rochester.edu}{S. Y. BenZvi} (University of Rochester), \href{mailto:rleane@mit.edu}{R. K. Leane} (Massachusetts Institute of Technology), \href{mailto:linden.70@osu.edu}{T. Linden} (The Ohio State University), \href{mailto:kenny.chunyu.ng}{K. C. Y. Ng} (Weizmann Institute of Science), \href{mailto:peter.33@osu.edu}{A. H. G. Peter} (The Ohio State University), \href{mailto:zhou.1877@buckeyemail.osu.edu}{\\B. Zhou} (The Ohio State University)\\
\noindent\textbf{Endorsers:} 
\href{mailto:dingus@lanl.gov}{B. L. Dingus} (Los Alamos National Lab),
\href{mailto:amalbert@lanl.gov}{A. Albert} (Los Alamos National Lab),
\href{mailto:klengel@umd.edu}{K. L. Engel} (University of Maryland), \href{mailto:hfleisch@mtu.edu}{H. Fleischhack} (Michigan Technological University), 
\href{mailto:green@liv.ac.uk}{T. Greenshaw} (University of Liverpool, UK), 
 \href{mailto:jpharding@lanl.gov}{J.~P.~Harding} (Los Alamos National Lab),
  \href{mailto:bhona@mtu.edu}{B.~Hona} (Michigan Tech),
 \href{mailto:petra@mtu.edu}{P.~Huentemeyer} (Michigan Tech),
\href{mailto:pierre.cristofari@gssi.it}{P. Cristofari} (Gran Sasso Science Institute),
\href{mailto:asandoval@fisica.unam.mx}{A. Sandoval} (Instituto de F\'isica, UNAM, M\'exico),
\href{mailto:harmscho@mpi-hd.mpg.de}{H. Schoorlemmer} (Max-Planck-Institut f\"ur Kerphysik, Heidelberg Germany),
\href{mailto:gilgamesh@upp.edu.mx}{Luis-Raya G.} (Universidad Polit\'ecnica de Pachuca, Pachuca Mexico),
\href{mailto:macj@cic.ipn.mx}{J. Mart\'inez-Castro} (Centro de Investigaci\'on en Computaci\'on-IPN, M\'exico),
\href{mailto:alberto@inaoep.mx}{A. Carrami\~nana} (INAOE, M\'exico),
\href{mailto:sako@icrr.u-tokyo.ac.jp}{T. Sako} (ICRR, University of Tokyo, Japan),
\href{mailto:miguel@psu.edu}{M.~A.~Mostaf\'a} (Pennsylvania State University),
\href{mailto:zepeda@fis.cinvestav.mx}{A. Zepeda} (Cinvestav),
\href{mailto:tollefson@pa.msu.ed}{K. Tollefson} (Michigan State University, USA),
\href{mailto:sdasso@iafe.uba.ar}{S. Dasso} (Instituto de Astronomía y Física del Espacio-IAFE, Argentina),
\href{mailto:cesar.alvarez@unach.mx}{C. Alvarez} (UNACH, M\'exico), 
\href{mailto:roberto.arceo@unach.mx}{R. Arceo} (UNACH, M\'exico), \href{mailto:karen.scm@gmail.com}{K. S. Caballero-Mora} (UNACH, M\'exico)
\\

\noindent \textbf{Abstract:}
The Sun is an excellent laboratory for astroparticle physics but remains poorly understood at GeV--TeV energies. Despite the immense relevance for both cosmic-ray propagation and dark matter searches, only in recent years has the Sun become a target for precision gamma-ray astronomy with the Fermi-LAT instrument. Among the most surprising results from the observations is a hard excess of GeV gamma-ray flux that strongly anti-correlates with solar activity, especially at the highest energies accessible to Fermi-LAT. Most of the observed properties of the gamma-ray emission cannot be explained by existing models of cosmic-ray interactions with the solar atmosphere. GeV--TeV gamma-ray observations of the Sun spanning an entire solar cycle would provide key insights into the origin of these gamma rays, and consequently improve our understanding of the Sun's environment as well as the foregrounds for new physics searches, such as dark matter. These can be complemented with new observations with neutrinos and cosmic rays. Together these observations make the Sun a new testing ground for particle physics in dynamic environments.

\normalsize

\clearpage

\setlength\itemsep{-0.1cm}

\section{Introduction}

The Sun is the most extensively studied star in a multitude of wavelengths. However, its complicated magnetic environment and its influence on cosmic rays at high energies remains a subject of open theoretical and observational investigations. It has long been of interest to the particle astrophysics community for its role in modulating the cosmic-ray flux in the solar system, and as a potential window to physics beyond the Standard Model. Cosmic rays entering the heliosphere are subject to propagation effects dominated by the solar wind and solar magnetic field. The cosmic-ray flux varies with solar activity throughout the 11-year cycle. In recent years, the time and energy dependence of the effect of solar magnetic fields has also been studied with the cosmic-ray shadow of the Sun \cite{PhysRevLett.120.031101,icecubecollaborationDetectionTemporalVariation2018,Astropaper}.

In neutrino astronomy $\gtrsim 1$ GeV, the Sun has been studied as a target for indirect detection of dark matter. Searches for annihilating dark matter in the Sun have been performed by Super-K \cite{Choi:2015ara}, IceCube \cite{2017EPJC...77..146A}, and ANTARES \cite{2016PhLB..759...69A,Adrian-Martinez:2016ujo} by looking for neutrino signatures of dark matter interactions. In the last few years, however, other messengers including gamma rays have become increasingly important due to,
(a) the increasing popularity of dark matter models with long-lived mediators,
(b) theoretical advances such as the identification of a neutrino floor for solar dark matter searches, (c) the first availability of high-statistics data on continuous emission from the Sun in the very-high-energy regime.

Long-exposure observations of the Sun extending into the GeV range only became a reality in the last decade. The launch of the Fermi-LAT instrument in 2008 enabled precision measurements of the gamma-ray Sun for the first time. Analysis of nine years of data collected by Fermi-LAT from the Sun revealed a very bright steady emission of gamma rays at energies above 100 GeV that contradicted all theoretical expectations \cite{2018arXiv180406846T}. This anomalous emission has become a new puzzle, the resolution of which would be a major step in our understanding of the Sun in an energy range that was not accessible before.

The problem calls for continued monitoring of the Sun in a broad range of energies going beyond what has already been observed by Fermi-LAT. In this regard, ground-based survey instruments, by virtue of their design, are capable of providing continuous, high-statistics data from the Sun. HAWC \cite{Astropaper} and ARGO-YBJ \cite{Bartoli:2019xvu} have already provided the first set of strong constraints on multi-TeV emission. Below we review the puzzling GeV spectrum of the Sun, identify the limitations of current gamma ray and neutrino searches, and discuss the imminent progress in the field with precision data on the Sun from the next generation of neutrino and gamma-ray observatories.    
\section{Observational Status and Open Questions}
\subsection{Cosmic-ray Induced Emission from the Sun}
The Sun is a steady emitter of multi-GeV gamma rays that originate from the Galactic cosmic rays interacting with its atmosphere. (Gamma-ray emission from particle acceleration during solar flares is restricted to $<4$ GeV.) There is a leptonic component of the emission that extends up to $20^{\circ}$ around the Sun. This halo has been robustly detected by Fermi-LAT and agrees well, at 0.1--10 GeV, with models of cosmic-ray electrons undergoing inverse Compton scattering off solar photons \cite{2008A&A...480..847O,Orlando:2006zs,Moskalenko:2006ta,Orlando:2013pza,Orlando:2017iyc}. On the other hand, a very bright emission of up to  200 GeV gamma rays from the solar disk is not understood \cite{Ng:2015gya,2018arXiv180305436L,2018arXiv180406846T}.

\begin{figure}[t!]
\includegraphics[width=1.05\linewidth]{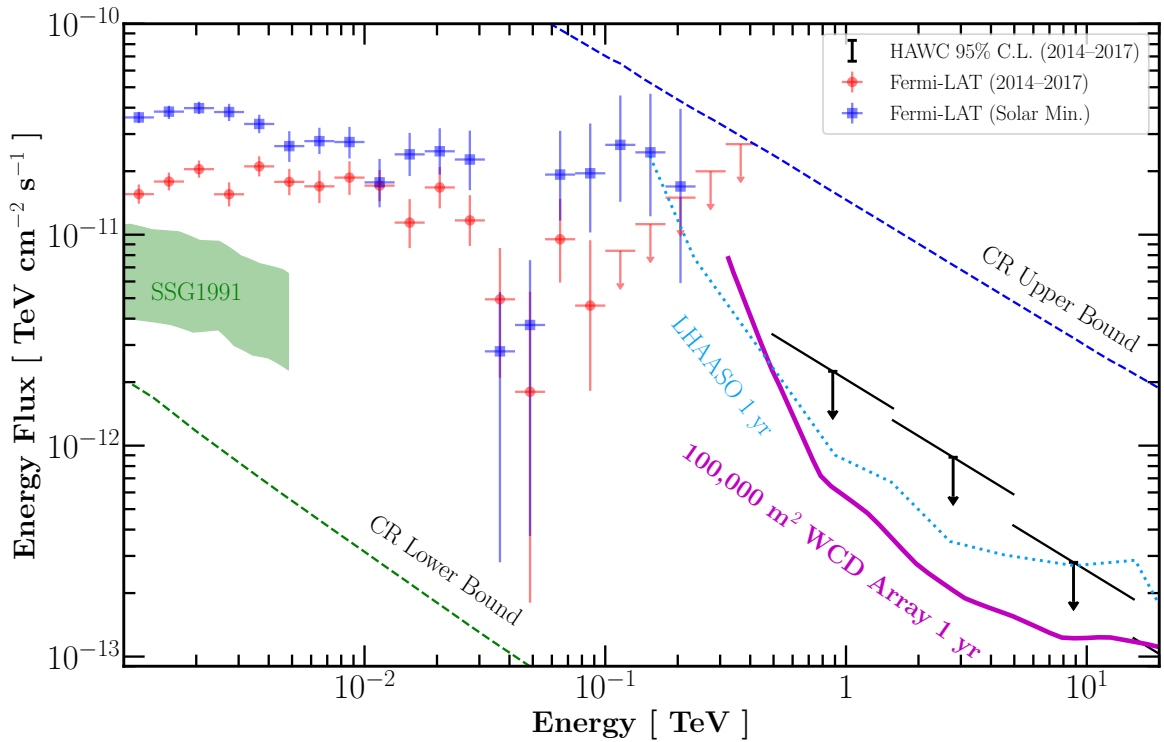}
  \caption{The observational status and future prospects of the solar gamma-ray spectrum. The red and blue data points show the Fermi-LAT measurements away from and during the solar minimum respectively. The nominal prediction from Ref. \cite{1991ApJ...382..652S} is shown as the green band. TeV limits from HAWC and the one-year LHAASO sensitivity show the possibility of constraining certain extrapolations of the GeV flux \cite{Astropaper}. (The ARGO-YBJ limits \cite{Bartoli:2019xvu} are just under the CR upper bound line.) The solid magenta line represents the potential gain in coverage from a next generation water Cherenkov detector (WCD) array in the Southern Hemisphere \cite{Albert:2019afb}.}
  \label{fig:obs}
\end{figure}

Hadronic cosmic rays entering the solar atmosphere can undergo magnetic mirroring, resulting in particle cascades that include a steady flux of gamma rays leaving the atmosphere without absorption. This model of gamma-ray production---proposed by Seckel, Stanev and Gaisser in 1991 \cite{1991ApJ...382..652S}---is in tension with observations in a number of ways. The measured flux between 0.1--10 GeV exceeds the theoretical predictions by a factor $\sim 6$. In fact, the flux measured at 100 GeV during the solar minimum very closely approaches the maximum possible flux (CR upper bound in Fig. \ref{fig:obs} and Ref. \cite{2018arXiv180305436L}) from cosmic-ray interactions. That implies almost every cosmic-ray hadron near the Sun is producing a visible gamma ray, which, is an efficiency not foreseen by any theoretical model. 

The overall flux follows a power-law spectrum with an index of $\sim$ E$^{-2.2}$, which is much harder than the expected E$^{-2.7}$ index (following the cosmic-ray spectrum). The spectrum also shows a mysterious dip around 40 GeV with no theoretical explanation \cite{2018arXiv180305436L,2018arXiv180406846T}.

Another intriguing observation is that the emission is   anti-correlated with solar activity, with the highest energy photons being produced during the solar minimum \cite{2018arXiv180305436L}. This time variation of the flux was also not predicted by theory. This is the first observed instance of the Sun exhibiting variability at such high energies. In addition, a study of resolved disk images by Linden et al. \cite{2018arXiv180305436L} found that the emission is non-uniform across the disk. In particular, it has a polar component that is roughly constant in time, and a second equatorial component that is brightest during the solar minimum. No single underlying mechanism has been found to explain these mysterious features. 

\begin{table}[t]
\begin{tabular}{lll}
\hline\hline
Feature & Theoretical Prediction & Observation \\
\hline
Flux (10 GeV) & $\sim 6 \times 10^{-12}$ TeV cm$^{-2}$ s$^{-1}$&  $\sim 2 \times 10^{-11}$ TeV cm$^{-2}$ s$^{-1}$ (solar min.)\\
Spectrum & $E^{-2.7}$ & $E^{-2.2}$\\
Dip & None & Significant dip in flux around 40 GeV\\
Time-dependence & None & Anti-correlated with Solar Activity\\
Morphology & Isotropic Point-like & Different Polar and Equatorial Components\\
\hline\hline
\end{tabular}
\caption{Differences between theoretical predictions of Ref. \cite{1991ApJ...382..652S} and nine-year observations of the Sun with the Fermi-LAT. \label{tab:problem}}
\end{table}


Table \ref{tab:problem} summarizes the discrepancies between theory and observation. Addressing these discrepancies calls for a detailed investigation of potential mechanisms that can enhance the flux of gamma rays, and answering the following questions,  
\begin{itemize}[noitemsep]

\item How far into the GeV-TeV range does the emission extend without a rigidity cutoff? What does that tell us about the spatial extent of the magnetic fields that confine the cosmic rays near the Sun?

\item Will the bright emission seen during the last solar minimum (2008--2010) repeat in the next cycle (2019--)? Is the amplitude of the modulation constant from one cycle to the next?

\item Is there a contribution to the emission from a new mechanism, and how do we use the spectral and spatial information to distinguish between multiple mechanisms of gamma-ray production? 

\end{itemize}

\subsection{Dark Matter Searches}
The search for particle dark matter is another motivation for precision studies of the Sun at GeV--TeV energies. Dark matter can be gravitationally captured in the Sun and settle in thermal equilibrium in the core following scatterings with solar nuclei. The dark matter in the core can annihilate to produce Standard Model (SM) particles, which may be detectable upon escaping the Sun and serve as a probe of dark matter-proton scattering rate \cite{Gould:1991hx,PhysRevLett.55.257,1995NuPhS..43..265E,2004PhRvD..69l3505L,2009arXiv0908.0899F,2009PhRvD..79j3532P,2011JCAP...09..029R,Danninger:2014xza,Choi:2015ara,Aartsen:2016zhm,2017JCAP...05..046W,Garani:2017jcj,Baum:2016oow,Ardid:2017lry}.
Neutrinos are the only SM particle that can escape when produced in the interior of the Sun, and so are the most sought after signature of dark matter annihilation in the Sun. In the well-motivated case that the dark matter first annihilates to long-lived mediators, neutrinos can be produced further away from the solar core, and so are less attenuated. Furthermore, other SM particles can escape the Sun to produce detectable gamma rays \cite{Meade:2009mu, 2010PhRvD..81g5004B,2010PhRvD..81a6002S, Bell:2011sn, Feng:2016ijc, 2017PhRvD..95l3016L, Arina:2017sng, Smolinsky:2017fvb}. The main foreground for these searches is the flux of neutrinos/gamma rays from hadronic cascades in the Sun'€™s atmosphere, which is an important complication that limits our sensitivity to dark matter signatures.

Gamma rays and neutrinos from dark matter have a distinct spectral and angular profile compared to the astrophysical foreground emission. Cosmic rays impinging on the back side of the Sun produce a steady flux of high-energy neutrinos. The flux of these solar atmospheric neutrinos has been estimated and, is potentially detectable with neutrino observatories on Earth \cite{2017PhRvD..96j3006N,2017JCAP...07..024A,Edsjo:2017kjk,2018APh....97...63M}. Neutrinos from solar dark matter annihilation would be correlated in direction with the center of the Sun, whereas the foreground neutrino emission would have a more extended profile with a dip towards the center \cite{Edsjo:2017kjk}. Once the neutrino detectors are sufficiently sensitive to the flux of solar atmospheric neutrinos, their sensitivity to dark matter searches will approach a soft sensitivity floor. The neutrino telescopes cannot distinguish between the solar atmospheric flux and dark matter due to their limited energy and angular resolution. 

On the other hand, the gamma-ray searches in TeV have yet to establish a sensitivity floor. Gamma rays from dark matter could be distinguished from cosmic-ray induced emission based on their spectrum and the time variation of the flux. Precision measurements of the above-mentioned astrophysical foregrounds and their accurate modeling are the major challenges that need to be addressed for future searches \cite{DMpaper}.

\subsection{Observational Challenges}
Figure \ref{fig:obs} illustrates the current status of observations and also shows the gaps in sensitivity and energy coverage that future studies need to fill to aid our understanding of the solar-disk spectrum. The Fermi-LAT has been able to measure photons up to 200 GeV, beyond which, satellite experiments have limited sensitivity. At higher energies, the strongest limits are available from the HAWC Observatory at energies 1--100 TeV for a search performed outside the solar minimum \cite{Astropaper,DMpaper}, and from ARGO-YBJ above 300 GeV \cite{Bartoli:2019xvu}. If the spectrum observed by Fermi during the last solar minimum continues into the next (2018--2020), then the prospects for a first TeV detection are promising given HAWC and the upcoming LHAASO sensitivity. However, at lower energies (300--800 GeV), neither HAWC nor ARGO-YBJ have sufficient sensitivity to exclude a simple E$^{-2.7}$ extrapolation of the spectrum measured by Fermi-LAT during any part of the solar cycle. Long exposure measurements in the energy range not covered by Fermi or current ground-based observatories is a key observational challenge for future. 

An important sensitivity benchmark is the theoretical lower bound \cite{Zhou:2016ljf} on the gamma-ray emission, which is still three orders of magnitude below the constraints provided by HAWC in the TeV range. This highlights another area of the parameter space where there is considerable room for improvement for future instruments. 

\section{Future Prospects for Solar Gamma-ray Studies}
\subsection{Experimental Frontier}
The next decade will see radical improvements in gamma-ray astronomy with a number of planned upgrades and new experiments on the horizon. While HAWC will continue to monitor the Sun for at least the next half of the solar cycle, LHAASO \cite{Zhen:2014zpa} beginning operations in 2020 will also be able to provide useful data with better sensitivity than HAWC. The Sun being a bright, moving source can only be efficiently probed using an all-sky survey instrument that is capable of day-time operations. Observation of the Sun is therefore beyond the reach of any Imaging Air Cherenkov Telescope (IACT) regardless of its sensitivity. While CTA \cite{Acharya:2017ttl} would be an excellent means of high-sensitivity pointed observations, it will not be able to probe the Sun due to intrinsic operational limitations. Only air-shower arrays  monitoring the whole sky can provide uninterrupted, high-statistics data from the Sun. We anticipate increasing importance of synoptic surveys to perform measurements of challenging extended sources that cannot be probed by both, satellites due to limited sensitivity, and, IACTs due to their limited field-of-view. A new gamma-ray survey observatory in the Southern Hemisphere would be a drastic improvement over the limitations of current arrays. We anticipate that the pursuit of gamma rays from the Sun would fit neatly in the goals of such an instrument.

\subsection{Physics Outcomes}
High-statistics gamma-ray measurements of the Sun and a potential resolution of the solar gamma-ray puzzle would be a significant step forward in our understanding of local cosmic-ray propagation in the Sun's dynamic environment \cite{Orlando:2017iyc,2018arXiv180305436L}. An observatory sensitive to cosmic rays in the GeV -- TeV range would be able to probe the coronal and interplanetary magnetic fields with long-term studies of the Sun shadow. However, Sun shadow measurements are prone to angular resolution limitations, and alone cannot provide insight into what fraction of the cosmic-ray flux is converted to gamma-ray emission. Sustained, long-term measurements of the Sun in gamma rays are therefore essential. 

The gamma-ray observations also complement searches for astrophysical neutrinos and dark matter from the Sun, resulting in increasingly stronger tests of new models. Dark matter-proton scattering limits from gamma-ray observations can outperform limits from direct-detection experiments by several orders of magnitude \cite{DMpaper}, and will be the world-leading probe of dark matter with precision measurements in the next decade. Moreover, a quantitative grasp of the solar magnetic fields can be used for accurate modeling of the trajectories and the deflection angles of charged particles from the Sun to the Earth. Progress on this front will allow us to perform directional searches for other secondary particles from the Sun's atmosphere, including electrons, positrons and neutrons, with fluxes comparable to those of gamma rays.  

Optical astronomy started with the Sun.  It is only fitting that in the emerging era of multimessenger astrophysics, the Sun should again hold a defining place.

\bibliographystyle{unsrt}
\bibliography{zbib}

\end{document}